\documentclass[aps,prl,twocolumn,
showpacs,
floatfix,nofootinbib,
superscriptaddress]{revtex4}
\pdfoutput=1

\usepackage{pstricks,epsfig,color}
\usepackage{psfrag}

\newcommand{\be}{\begin{equation}}
\newcommand{\ee}{\end{equation}}
\newcommand{\bear}{\begin{eqnarray}}
\newcommand{\eear}{\end{eqnarray}} \newcommand{\ba}{\begin{array}}
\newcommand{\ea}{\end{array}}
\newcommand{\lae}{\begin{array}{c}\,\sim\vspace{-1.7em}\\< 
\end{array}}

\def\beq{\begin{equation}}
\def\eeq#1{\label{#1}\end{equation}}
\def\eeqn{\end{equation}}
\def\eeq{\end{equation}}
\def\beqa{\begin{eqnarray}}
\def\eeqa#1{\label{#1}\end{eqnarray}}
\def\eeqan{\end{eqnarray}}

\def\to{\rightarrow}

\newcommand\iden{\leavevmode\hbox{\small1\normalsize\kern-.33em1}}

\def\W3{W_H^3}

\begin{document}

\title{Is the LHC Observing the Pseudo-scalar State of a Two-Higgs Doublet Model ?}
\author{Gustavo Burdman
%\footnote{E-mail: burdman@if.usp.br}
}
\author{
Carlos E.~F.~Haluch
%\footnote{E-mail: lascio@fma.if.usp.br}
}
\affiliation{Instituto de F\'isica, Universidade de S\~ao Paulo, 
S\~ao Paulo, Brazil}
\author{
Ricardo D.~Matheus
%\footnote{E-mail: lascio@fma.if.usp.br}
}
\affiliation{
Instituto de F\'isica Te\'orica, Universidade Estadual  Paulista, S\~ao Paulo, Brazil}
\vspace*{0.3cm}
%\date{\today}

%\vspace*{0.4in}

\begin{abstract}
%\vspace*{0.2in}
The ATLAS and CMS collaborations have recently shown data suggesting
the presence of a Higgs boson in the vicinity of $125~$GeV. 
We show that a two-Higgs doublet model spectrum, with the pseudo-scalar
state being the lightest, could be responsible for the diphoton signal
events. In this model, the other scalars are considerably heavier and
are not excluded by the current LHC data. If this assumption is
correct, future LHC data should show a strengthening of the
$\gamma\gamma$ signal, while the signals in the $ZZ^{(*)}\to 4\ell $ and $WW^{(*)}\to
  2\ell 2\nu$ channels should diminish and eventually disappear, due
  to the absence of diboson tree-level couplings of the CP-odd state.
The heavier CP-even neutral scalars can now decay into channels
involving the CP-odd light scalar which, together with their larger
masses, allow them to avoid the existing bounds on Higgs searches.
We suggest  additional signals to confirm  this scenario at the LHC,
in the decay channels of the heavier scalars into $AA$ and $AZ$. 
Finally, this  inverted two-Higgs doublet spectrum is characteristic in models
where fermion condensation leads to electroweak symmetry breaking. We
show that in  these theories it is possible to obtain the observed diphoton signal
at or somewhat above of the prediction for the standard model Higgs
for the typical values of the parameters predicted.

\end{abstract}
%}

\maketitle

\section{Introduction}  Recently the ATLAS and CMS collaborations have reported important
exclusions in the Higgs mass with about $5$ fb$^{-1}$ of accumulated
luminosity~\cite{atlas,cms}. 
However, both experiments observe excess signal events
at low masses.  The most significant of these is in the diphoton channel
$h\to\gamma\gamma$ channel, which would point to a Higgs mass
of $m_h\simeq 126~$GeV.  For ATLAS~\cite{atlasgg}  the  local significance of this excess is of
$2.8~\sigma$, whereas for CMS~\cite{cmsgg} it is about
$3.0~\sigma$. 
ATLAS~\cite{atlaszz}  and CMS~\cite{cmszz} also observe modest excesses
in the $h\to ZZ^{(*)}\to
4\ell$ channel with local significance not musch above $2.0~\sigma$, as well as
CMS~\cite{cmshjj} has also searched for an enhanced $\gamma\gamma j j$ signal coming
from vector boson fusion (VBF), which could be associated with fermiophobic
Higgs models. With cuts that significantly reduce the gluon fusion
contribution, CMS has found a $2.7~\sigma$ excess. Finally,
ATLAS~\cite{atlashjj} has
also search for similar signals, which are enhanced by a cut in the
transverse momentum of the diphoton system. They found an excess of
about $3~\sigma$. 
Thus, taken all combined, both experiments appear to coincide in the
existence of a diphoton excess somewhere around $(124-126)~GeV$, while
differing on the invariant mass of the less significant excesses
observed in $ZZ^{(*)}\to4\ell$, as well as in $WW^{(*)}\to\ell^+\nu\ell^-\bar\nu$.

In this paper we consider the possibility that the excess in the
$\gamma\gamma$ channel is real, but that it is caused by the pseudo-scalar
state $A$ in a two-Higgs doublet model (THDM). We assume that the
spectrum of the THDM is inverted with respect to what is  usually
considered, for instance in the Minimal Supersymmetric Standard Model
(MSSM)~\cite{mssm}:
 $A$ is the lightest state, with
$(h,H,H^\pm)$ much heavier and with small splittings among
them~\cite{maltoni,sher}. 
In considering this possibility, we must address the VBF-enriched
samples in Refs.~\cite{cmshjj} and \cite{atlashjj}, which find excesses
in regions that favor VBF and disfavor gluon fusion. However, the
excesses are still small ($3~\sigma$ and $2.7~\sigma$ for CMS and
ATLAS respectively), and the techniques used new. In any case, the pure gluon fusion interpretation is still
compatible with both the ATLAS and CMS  data at about the $3~\sigma$ level~\cite{strumia}. So more data is
needed. 

The inverted spectrum can naturally  appear in  a generic THDM, 
just by choosing the right set of parameters in the potential, even
after demanding tree-level unitarity and stability, as well as the
correct minimum for electroweak symmetry
breaking~\cite{brancosher,maltoni}.  
On the other hand, this spectrum of the THDM is typical  in models of where un-confined fermions
condense to break the electroweak symmetry dynamically~\cite{fgthdm},
due to the existence of an approximate Peccei-Quinn symmetry which
keeps the pseudo-scalar state light in comparison t the rest of the scalars. 

 Whatever the origin of this scalar spectrum, the first signal of
it would be the observation of $A\to \gamma\gamma$ if $m_A<130~$GeV,
just as is the case of the SM Higgs. The production cross section
times branching ratio for $gg\to A\to\gamma\gamma$ need not be 
the same as that of the SM Higgs. In fact the ATLAS signal is somewhat
larger that the SM prediction for a Higgs of $\simeq 125~$GeV. 
On the other hand, since $A$ has no
tree-level couplings to $ZZ$ and $WW$, these channels would not be
observed for the lightest mass peak. More importantly, the current
excesses in the VBF channels studied by ATLAS and CMS should
disappear. 
If this is the case, other channels will have to be studied to confirm
the THDM explanation.
In what follows, we specify the parameter space of this scenario as
well as its predictions for future LHC data samples. We also point out
the strategy for finding the other states  that would
confirm the THDM hypothesis. Finally, we speculate on the possible
dynamical origin of this peculiar THDM spectrum and its relation to
theories of electroweak symmetry breaking (EWSB).

\section{Predictions in the inverted THDM} 
In order to specify the phenomenology of the THDM we must define the
scheme of fermion couplings. There are four possible choices that
would avoid flavor changing neutral currents (FCNCs) at tree level:
the so called type I, type II, lepton-specific and flipped
schemes~\cite{barger,brancosher}, depending on the choice of doublet
responsible of the masses of right-handed fermions. The couplings of
the scalar sector are determined by $\tan\beta\equiv v_2/v_1$, the
ratio of vacuum expectation values of the two doublets, and the mixing
angle $\alpha$ between the neutral CP-even scalars. However, the couplings of
$A$ to fermions only depend on $\beta$. For
instance, the
couplings of $A$ to all up-type quarks in all four schemes is
$\cot\beta$. On the other hand, its couplings to down-type, go like
$\cot\beta$ in the type~I and lepton-specific case, whereas they go
like $\tan\beta$ for the type~II and flipped schemes. Finally, the
 couplings of charged leptons to $A$ in type~I and flipped scenarios
 go like $-\cot\beta$, while the same couplings in type~II and
 lepton-specific schemes go like $\tan\beta$. 

We are interested in calculating the $\sigma\times BR (gg\to A\to
\gamma\gamma)$ in the inverted THDM.  The production cross section 
for $gg\to A$ is given by~\cite{spira}
\begin{equation} 
\sigma_A = \frac{9}{4}\,\frac{\left|\cot\beta\,
    I_A(\tau_t) + \xi(\beta)\,I_A(\tau_b)\right|^2}{\left|I_S(\tau_t)\right|^2}\, \sigma_h^{\rm SM}~,
\label{aprod}
\end{equation} 
where $\sigma_h^{\rm SM}$ is the SM $gg\to h$ production cross
section, 
the functions  in (\ref{aprod})  are given by:
\begin{eqnarray}
I_A(\tau)  &=&\frac{f(\tau)}{\tau} ~, \label{ia}\\
I_S(\tau) &=&\frac{3}{2}\,\left[\tau+(\tau-1)f(\tau)\right]/\tau^2~,
\label{is}
\end{eqnarray}
and 
\begin{equation}
f(\tau) = \left\{
\begin{array}{lr}
\arcsin^2\sqrt{\tau}~, &  \tau\leq1 \\
-\frac{1}{4}\,\left[\log\frac{1+\sqrt{1-\tau^{-1}}}{1-\sqrt{1-\tau^{-1}}}
  -i\pi\right]^2~, & \tau>1
\end{array}
\right.
\end{equation}
expressed in terms of  the variable $\tau_f\equiv
m_{A,h}^2/4m_f^2$. The factor $\xi(\beta)$ in (\ref{aprod}) depends on
the choice of THDM scheme: for type~I and lepton specific models,
$\xi(\beta)=-\cot\beta$, while for type~II and flipped cases we have $\xi(\beta)=\tan\beta$.
For moderate values of $\tan\beta$ the top quark contribution
dominates the $ggA$ vertex. The b quark term is important for large
$\tan\beta$ in the type~II and flipped schemes. 

Similarly, we can compute the $A\to \gamma\gamma$ decay width
normalized by the SM $h\to \gamma\gamma$  width, obtaining
\begin{eqnarray}
&&\frac{\Gamma(A\to\gamma\gamma)}{\Gamma^{\rm SM}(h\to\gamma\gamma)} =\nonumber\\
&&4\,\frac{\left|N_c q_U^2 \cot\beta I_A(\tau_t) + N_c q_D^2\xi(\beta)
    I_A(\tau_b) \right|^2}{\left|N_c(4/3)(q_U^2 I_S(\tau_t) + q_D^2 I_S(\tau_b))
    I_W(\tau_W)\right|^2}~,
\label{adecay} 
\end{eqnarray}
where $q_{U,D}$ are the charges of up and down type quarks and the $W$
contribution to the SM Higgs decay to diphotons is accounted for  by the
function
\begin{equation}
I_W(\tau) = -\left[2\tau^2 + 3\tau + 3(2\tau-1)f(\tau)\right]/\tau^2~.
\label{iw} 
\end{equation} 

We are now ready to compute the $\sigma\times BR(gg\to
A\to\gamma\gamma)$ in the inverted THDM normalized by the analogous SM
Higgs process. We have not included QCD corrections to the quark
contributions since they largely cancel in the ratio.  
 \begin{figure}
\includegraphics[scale=0.6]{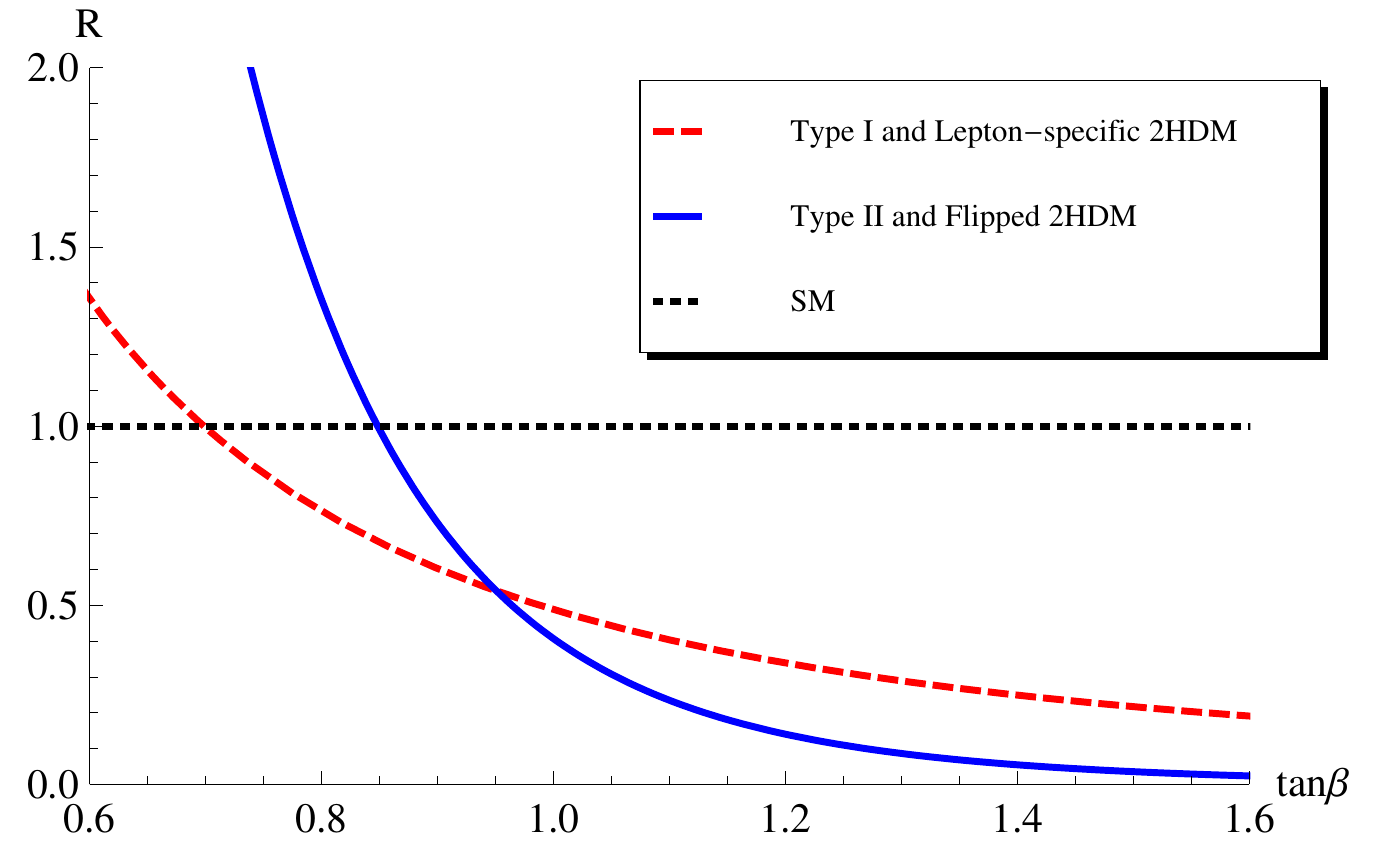}
\caption{The ratio $R=\sigma\times BR (gg\to
A\to\gamma\gamma)/\sigma\times
BR(gg\to h\to\gamma\gamma)$ vs. $\tan\beta$. The dashed line corresponds
to the type~I and lepton-specific schemes, with the solid curve being
for the type~II and flipped cases. The horizontal dotted line
corresponds to $\sigma\times BR(gg\to A\to\gamma\gamma)$ equal to the
prediction for this process mediated by the SM Higgs.}
\label{fig:1}
\end{figure}
 In Figure~\ref{fig:1} we plot the ratio  $R=\sigma\times BR (gg\to
A\to\gamma\gamma)/\sigma\times
BR(gg\to h\to\gamma\gamma)$ vs. $\tan\beta$. The dashed line corresponds
to the type~I and lepton-specific schemes, with the solid curve being
for the type~II and flipped cases. 
We see that excesses over the SM prediction for $gg\to
h\to\gamma\gamma$ can only be obtained for  $\tan\beta<0.9$ or
so. Larger values of $\tan\beta$ would imply $R<1$, in contradiction
with the ATLAS data, as long as we interpret it as purely coming from
signal (i.e. not aided by a significant upward fluctuation). 
The top Yukawa coupling only becomes non-perturbative for  $\tan\beta<0.3$,
so these values are safe~\cite{brancosher}.  
Thus, we see that the inverted THDM can explain the diphoton signal at
the LHC for these small values of $\tan\beta$. 

\begin{figure}
\includegraphics[scale=0.8]{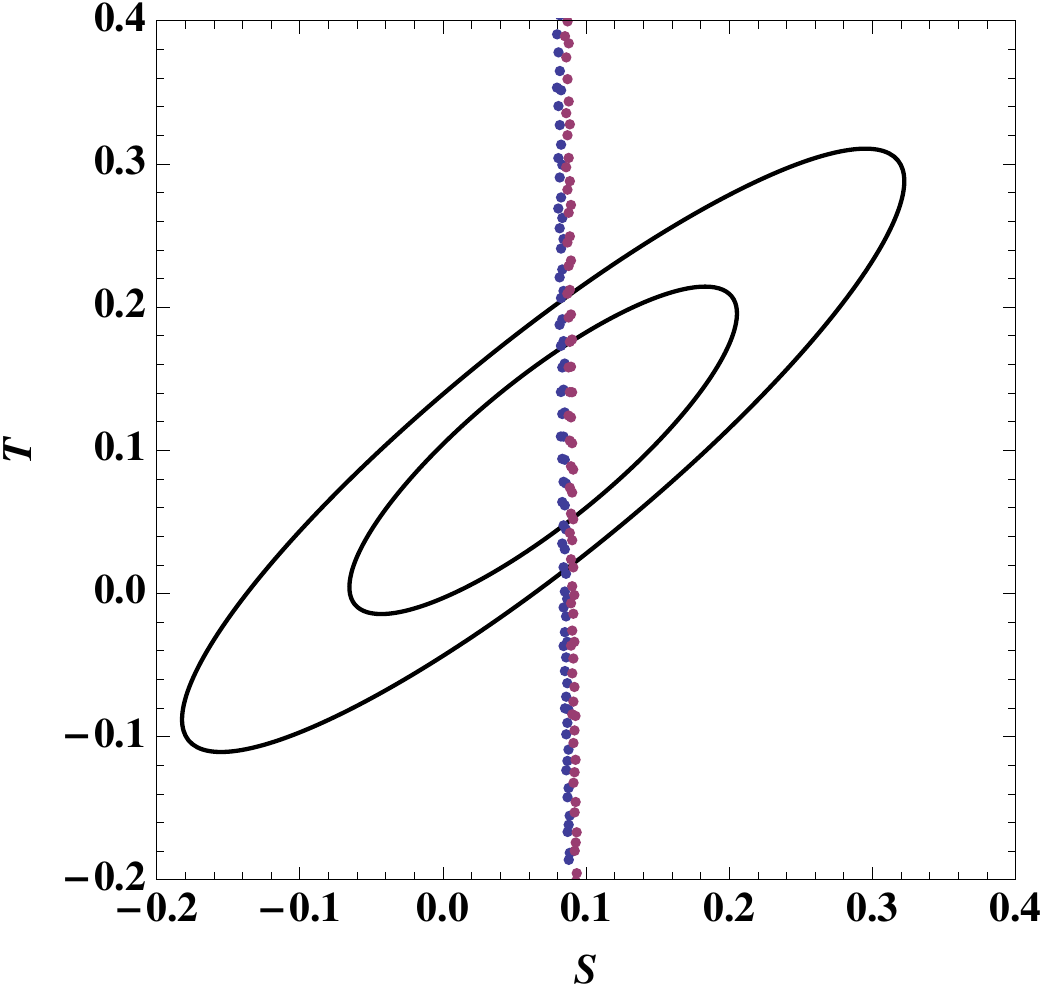}
\caption{Electroweak precision bounds on the inverted THDM. We tale 
$m_A=125~$GeV, and $m_h=(500,550)$, with $m_H$ and $m_{H^\pm}$
varying in the range $(550-650)~$GeV. A scan over this parameter space
results in the points in the figure. The ones outside the
$95\%$~C.L. contour come from larger values of $|m_H-m_{H^\pm}|$,
which result in large custodial breaking. We take $m_h^{\rm ref.}=117~$GeV~\cite{pdg}.
}
\label{fig:st}
\end{figure}

This region of
parameter space is allowed by all other data. In addition to the
absence of FCNC at tree level, the loop contributions to FCNC
processes are safely below bounds given that the charged states of the
model, $H^\pm$, are assumed to have masses well in excess of the $95\%~$C.L. bound
$m_{H^\pm}>316~$ GeV which is mostly driven by $b\to
s\gamma$~\cite{flavorbounds}.  It is also generally compatible with
precision electroweak constraints. In Figure~\ref{fig:st} we show the
contributions to the $S$ and $T$ parameters the inverted THDM. In
particular we take  $m_A=125~GeV$, and a scalar spectrum that is heavy
enough not to have been seen at the LHC~\cite{inprepa} so far. 
For the two choices $m_h=500$~GeV and $m_h=550$~GeV, we vary $m_H$ and $m_{H^\pm}$ in the range
$(550-650)~$GeV. The points represent the results of scanning over
this  range. We also show the $68\%$ and $98\%$ ~C.L. intervals
obtained from a fit of electroweak data as performed in Ref.~\cite{pdg}
using  $m_h^{\rm ref.}=117~$GeV.
We can see from Figure~\ref{fig:st} that there are
many solutions within the inverted THDM spectrum that are compatible
with electroweak bounds. The points falling out of the allow region
are those with large values of $|m_H-m_{H^\pm}|$, which represent a breaking
of custodial symmetry. As long as this mass difference is not large
compared to $m_W$, the inverted THDM spectrum is within the allowed
values of the electroweak parameters $S$ and $T$. 

\begin{figure}
\includegraphics[scale=0.6]{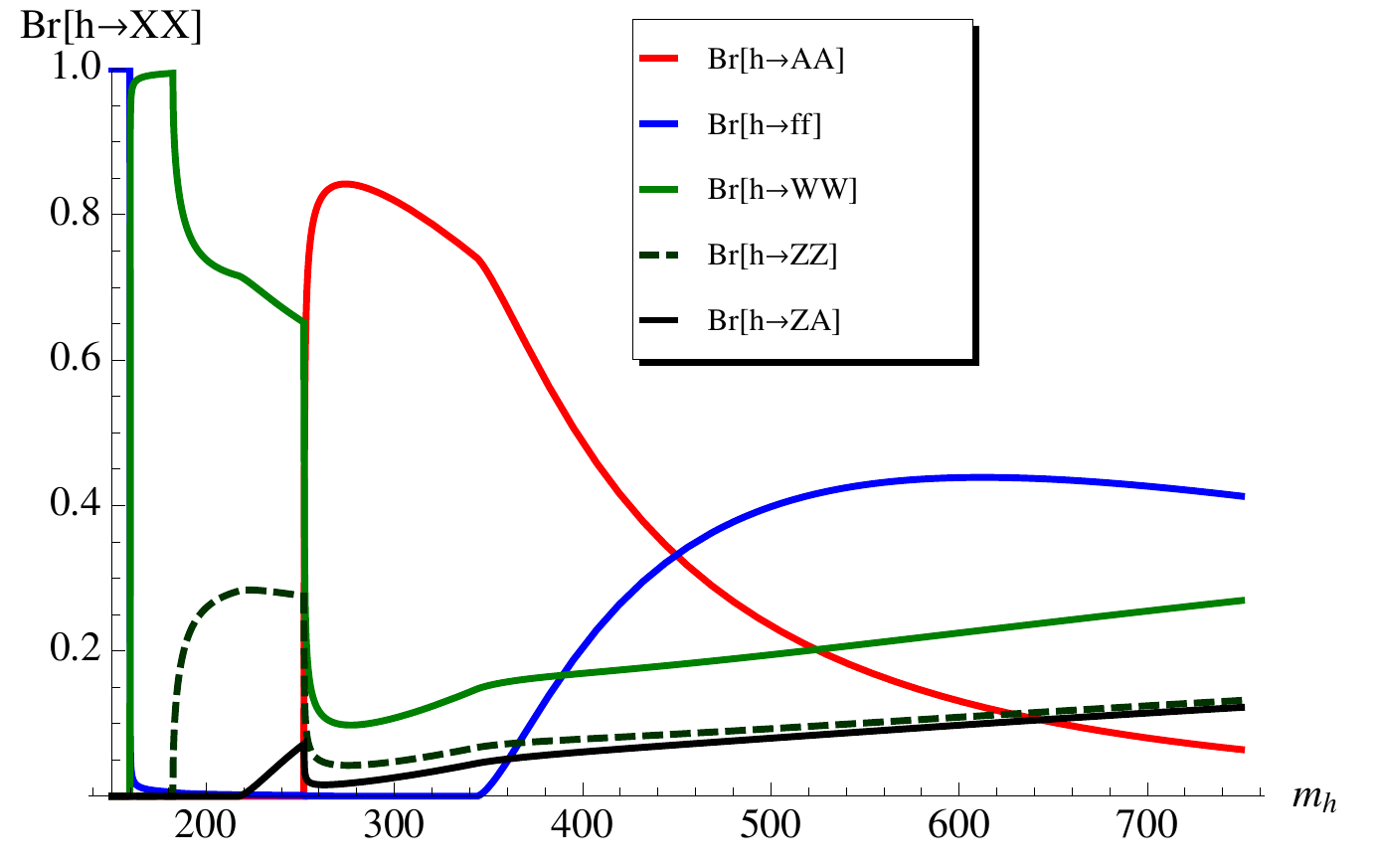}
\caption{Branching ratios for the lightest CP-even $h$ vs. $m_h$, for
  $\tan\beta=0.8$,  $\alpha=-0.006$, and $m_A=126~$GeV. Here $f$
  refers to all SM fermions, and the couplings to them are type~II.
}
\label{fig:2}
\end{figure}

The bounds on Higgs searches from ATLAS and CMS suggest that the rest
of the scalar spectrum of the THDM is quite heavy.  A detailed
study of the exclusion is left for  Ref.~\cite{inprepa}. 
But we can see that 
the neutral states, $h$ and $H$, are not excluded by the SM Higgs
bounds obtained using the standard channels, due to a combination of
them being heavy plus the fact that other decay channels are
open for their decays. In particular, the decay channels 
\begin{equation}
(h,H)\longrightarrow AA\qquad  (h,H)\longrightarrow AZ \label{newdecay}
\end{equation}
that are competitive with $(h,H)\longrightarrow WW$ and
$(h,H)\longrightarrow ZZ$, the channels that drive the bounds on
large SM Higgs masses. In order to illustrate this we show in Figure~\ref{fig:2} the
branching fractions of the lightest CP-even neutral scalar $h$ for the
type~II case..
The branching ratios  are computed for $m_A=126~$GeV and  $\tan\beta = 0.8$, which would
correspond to a diphoton signal at about the same level of the SM
Higgs, as can be seen  in Figure~\ref{fig:1}.  We used a negligibly
small value of the 
mixing angle $\alpha=-0.006$ which comes from the typical parameter
space studied here and also is typical in the models presented in the
next section. This  results in an important coupling of $h$ to the top
quark, making $t\bar t$  an important decay channel above 
threshold. More importantly,  we see that the $ZZ$ and $WW$ channels
used by ATLAS and CMS to put bounds on the mass of the Higgs now have
to compete with the $AA$ and $AZ$ channels. In fact, the LHC bounds on Higgs
searches do not apply to the THDM $h$ once its mass is above $AA$
threshold or about $250~$GeV. The $ZZ$ only becomes more important
than the $AA$ channel for rather large masses. 

\begin{figure}
\includegraphics[scale=0.6]{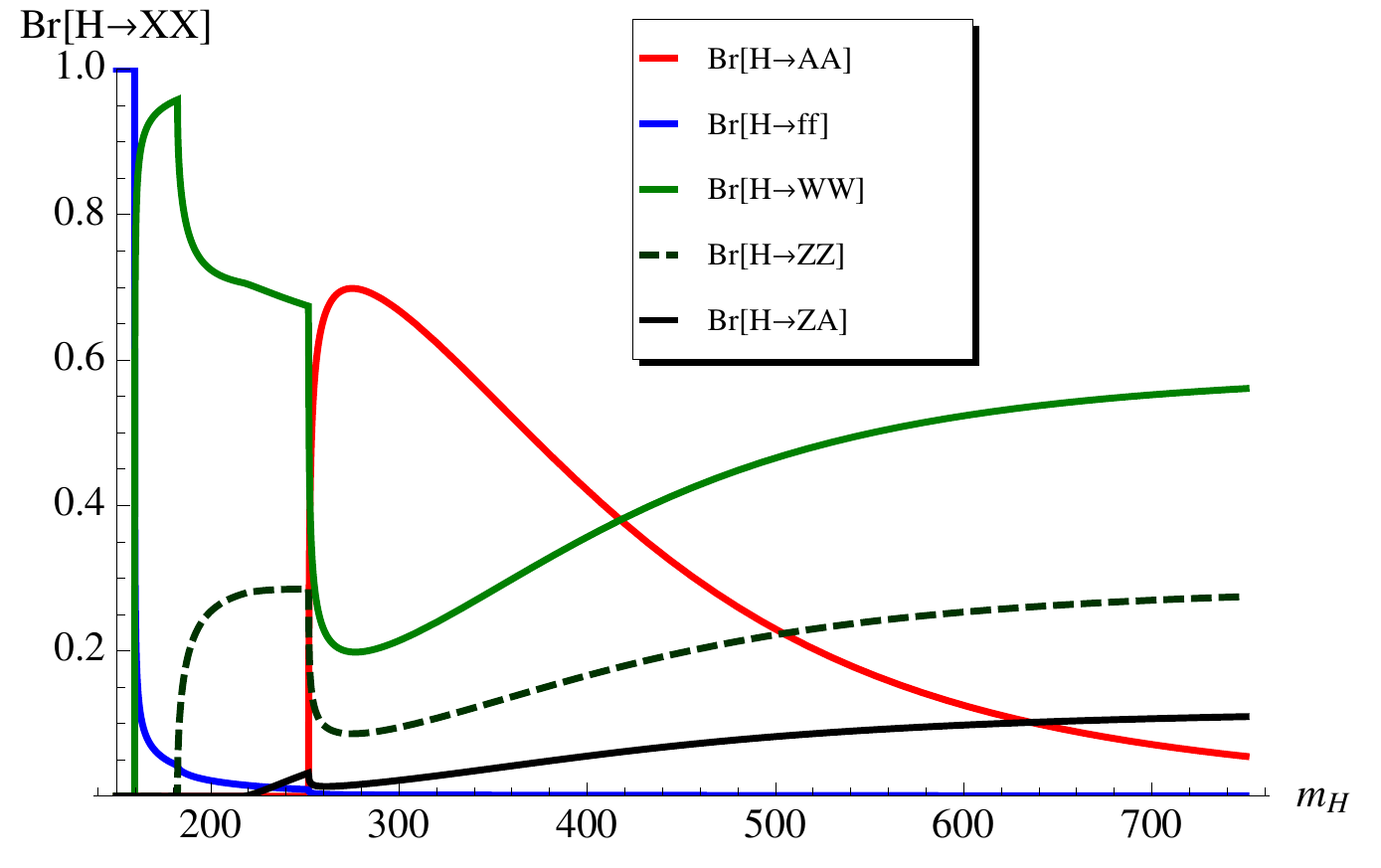}
\caption{Branching ratios for the heaviest CP-even,  $H$,  vs. $m_H$, for
  $\tan\beta=0.8$, $\alpha=-0.006$, and $m_A=126~$GeV. Here $f$
  refers to all SM fermions, and the couplings to them are type~II.
}
\label{fig:3}
\end{figure}

Similarly, we plot the the branching fractions for the decays of the
heavier CP-even state, $H$, in Figure~\ref{fig:3}. 
Thus, we see that the search for the neutral states in the inverted
THDM should include, in addition to the $ZZ$ and $WW$ channels, the
$(h,H)\to AA\to 4b's$ and $(h,H) \to AZ\to b\bar
b\ell^+\ell^-$. In both cases the $A$ should be reconstructed to have
the same mass as the one in the $\gamma\gamma$ channel, which would
provide a nice confirmation of the model~\cite{inprepa}.  

The spectrum of the inverted THDM presented here, with a light
pseudo-scalar with $m_A\sim 125~$GeV, and heavy masses for the scalars
$(h,H,H^\pm)$ in the $(500=600)~$GeV range, is compatible with
tree-level unitarity. To check this we test the parameters of the
model by calculating the scattering matrix of scalar interactions. We
require that the eigenvalues of this matrix be smaller than $16\pi$,
corresponding to saturation. All the points in the parameter space
considered here satisfy this constraint~\cite{tlunitarity}.
On the other hand, the large masses for the scalars point to the
existence of a strongly coupled sector wich should appear not too far above the
scalar masses. Thus, this scenario for the scalar spectrum of the
THDM should be accompanied by new states not too far above the scalar
masses, such as new gauge bosons and/or new fermions. We will present
such an example in the next section.

\section{A Model of the Inverted THDM}  Although in principle the
inverted THDM spectrum can always be considered as a possibility, it
is interesting to ask whether  such spectrum can be obtained 
dynamically.  The typical MSSM scalar spectrum requires a light
scalar, so the pseudo-scalar is rather heavy, or much lighter as in
the NMSSM~\cite{nmssm}. A techni-pion in techni-color models 
could be light~\cite{sekhar}, but the rest of the THDM is missing.  
 In Ref.~\cite{fgthdm} it was shown that this inverted
THDM is precisely obtained in theories where  the condensation of a
fermion sector leads to EWSB.  In
these models the new chiral fermions feel a strong interaction at the
few TeV scale, condensing and breaking the electroweak symmetry. Since
both the up-type and down-type right-handed
fermions condense with the left-handed doublet, the resulting scalar
spectrum at low energies is that of the THDM. Furthermore, the
presence of a Peccei-Quinn symmetry in the fermion theory, only broken
by the new interaction's instanton effects, guarantees that the CP-odd
state $A$ is the lightest state of the scalar spectrum.  The rest of
the spectrum, given by $(h,H,H^\pm)$, is much heavier as is expected
from the condensation of fermions with a cutoff scale in the multi-TeV
region. The scalars are also  somewhat degenerate giving this inverted
THDM a very interesting phenomenology at the LHC. The bounds from
electroweak precision measurements vary somewhat in the presence of the
new fermions, but it is still possible to have agreement with them,
as shown in Ref.~\cite{fgthdm}.

\begin{figure}
\includegraphics[scale=0.6]{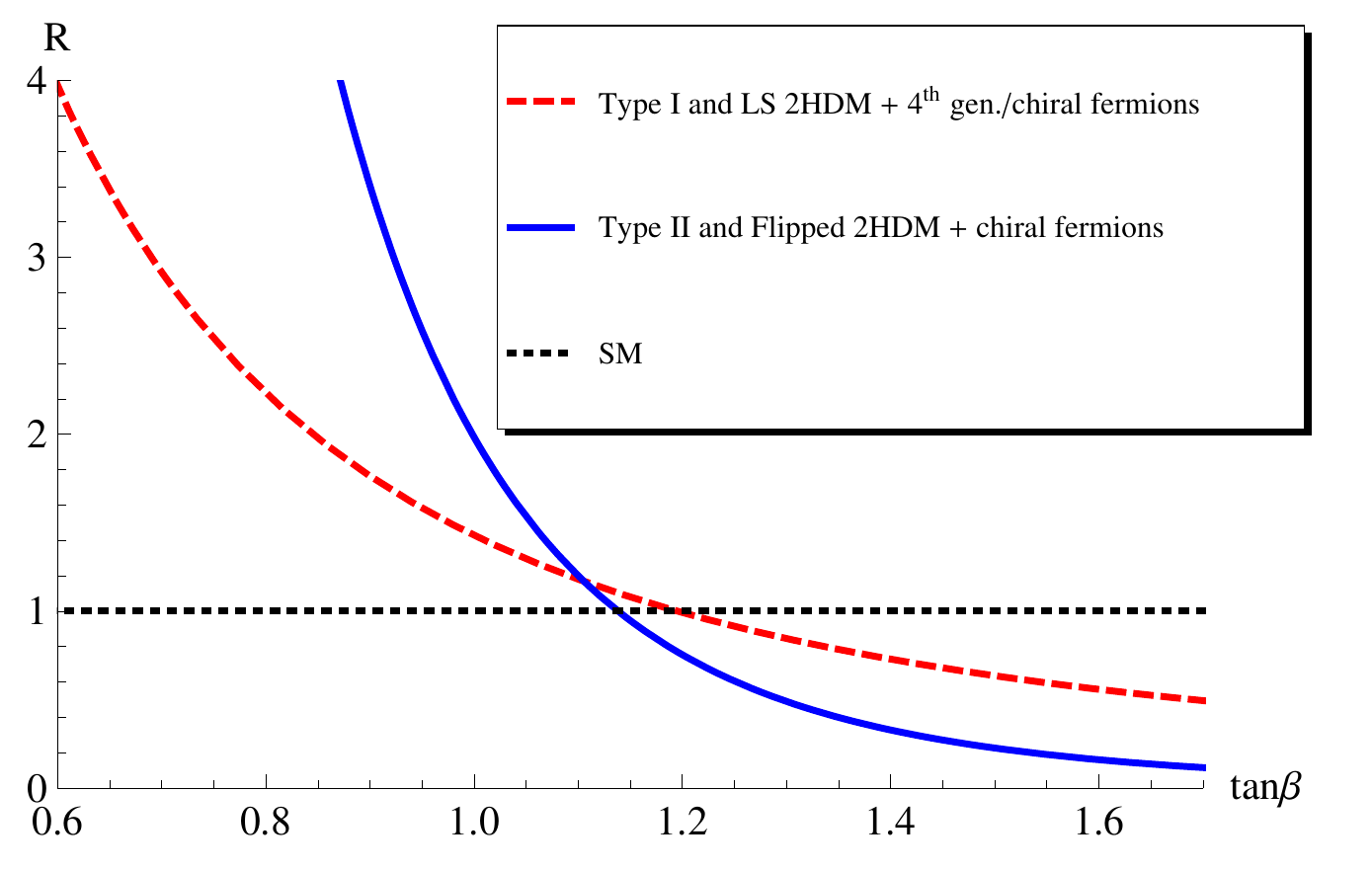}
\caption{The ratio $R=\sigma\times BR (gg\to
A\to\gamma\gamma)/\sigma\times
BR(gg\to h\to\gamma\gamma)$ vs. $\tan\beta$, for $m_A=126~$GeV, in a model with colorless
chiral fermions. The line of for the type~II scheme. 
The horizontal dashed line
corresponds to $\sigma\times BR(gg\to A\to\gamma\gamma)$ equal to the
prediction for this process mediated by the SM Higgs.}
\label{fig:4}
\end{figure}

The simplest model for the condensing fermions is to assume that they
are quarks  belonging to a fourth generation~\cite{fg1}. However, 
this assumption together with the hypothesis that
the diphoton signal comes from the pseudo-scalar $A$ is in great
tension with  the recent LHC data.  In order to see this, we first consider  the
contributions of the fourth generation quarks to the $ggA$ vertex. 
The dynamics of the condensation naturally selects the type~II
scheme for the fourth generation~\cite{fgthdm}, so it is natural to adopt it for all four
generations.  With the addition of the fourth-generation quarks, 
now Eq.~(\ref{aprod}) for the $gg\to A$ production cross section reads
\begin{eqnarray} 
\frac{\sigma_A}{\sigma_h^{\rm SM}} \, \left|I_S(\tau_t)\right|^2 & =&
\frac{9}{4}\,\left|\cot\beta 
    I_A(\tau_t) + \tan\beta I_A(\tau_b) \right.\nonumber\\
& &\left. + \cot\beta  I_A(\tau_{t'})
    + \tan\beta I_A(\tau_{b'})
\right|^2~,
\label{aprodfg}
\end{eqnarray}
where the last two terms are the contributions of the $t'$ and $b'$
fourth-generation quarks, respectively. 
As pointed out in Ref.~\cite{fgthdm}, the condensation models typically select
$\tan\beta\simeq 1$. Thus, we see that the production cross section is
greatly enhanced, by a factor of about 9 at leading order. 
On the other hand, and unlike the case for the Higgs in the presence
of a fourth generation~\cite{he}, the $A\to \gamma \gamma$ is not
suppressed in the presence of the fourth generation since there is no
W contribution against which to cancel . To the numerator in the expression
(\ref{adecay}), we must now add 
\begin{equation}
N_c\, q_U^2 \cot\beta\, I_A(\tau_{t'}) + N_c\, q_D^2\tan\beta\,
    I_A(\tau_{b'})~.
\label{fgadd}
\end{equation} 
As a result the $A\to\gamma\gamma$
branching ratio is also  enhanced.   
All in all, the $\sigma\times
BR(gg\to A\to\gamma\gamma)$ in the fourth generation model is larger
than the SM analogous rate by a factor of 10 for $\tan\beta=1$ and 
$m_A=126~$GeV, which is excluded by the LHC data. 
Although not motivated by the condensation models, we could consider
type~I THDM with a fourth generation. The resulting ratio of $\sigma\times BR (gg\to
A\to\gamma\gamma)$ to the corresponding SM prediction for the Higgs is
plotted as the dashed line in Fig.~\ref{fig:4}. We see that for this
case, it is possible to generate rates consistent with or slightly above
the SM rates for a Higgs of the same mass, as long as $\tan\beta\lae 1.1$.

Another possibility for generating the inverted THDM spectrum is to just
consider that the condensing fermions, although chiral, do not carry
color~\cite{fgthdm}.  In this case the strong interaction responsible
for fermion condensation is not related to $SU(3)_c$. With  this
assumption, the new heavy fermions do not contribute to the $gg A$
vertex, leaving this to be just as in Eq.~(\ref{aprod}) for the three
generation case. On the other hand, the fact that the new fermion
condensates break the electroweak symmetry, implies that they must have
couplings to the photon, and therefore they will contribute to the $A
\gamma\gamma$ vertex. If for concreteness  we assume that the exotic
fermions have the same charges as the up and down quarks, and that the
new strong interaction requires that there be $N_f$ copies of them
(e.g. the new interaction is $SU(N_f)$), we can make a definite
prediction for the new fermion contributions to the
$\Gamma(A\to\gamma\gamma)$ by just using (\ref{adecay}) supplemented by
(\ref{fgadd}) with $N_f$ substituting $N_c$. 
As an example, in the solid line of
Figure~\ref{fig:4} we plot  the ratio $R$ for a model with colorless
chiral fermions including a doublet and two singlets with the same
hypercharges as the fourth generation quarks, and for
$N_f=3$. Although the new
fermions are typically  heavy ($m_f\simeq 600~$GeV),  the results depend
little on the value of their mass as long as they are significantly
heavier than the top quark.  We see
that, although the curve is somewhat shifted upwards with respect to
the three-generation THDM case of Figure~\ref{fig:1}, it is still
possible to obtain cross sections for the $\gamma\gamma$ channel that
are  of order of the SM Higgs ones, or even somewhat larger as long as
$\tan\beta\simeq O(1)$, which is the value selected by the dynamics of
the fermion condensation models.

\section{Conclusions} The ATLAS and CMS data are beginning to probe
the scalar sector of the SM, the least known sector of the theory.
It is important that we consider alternatives to the one doublet case,
and that we study the resulting phenomenology in light of  the coming
increased  data samples.    
In this letter we have studied the
possibility that the diphoton signal observed in ATLAS and CMS is due to the
pseudo-scalar, CP-odd, state  $A$ of a THDM where it is the lightest
scalar in the spectrum. If this were the case, no $ZZ$ and $WW$ signal
should be confirmed at the diphoton invariant mass, $m_A\simeq
125~$GeV.  Furthermore, the  VBF-enriched channels of diphoton
production, studied by both
collaborations with different methods should, be accounted for only
from the gluon fusion process. These channels are new and the excesses
found are still small. But if they are confirmed with more data they
would put severe constraints to their interpretation as coming purely
from gluon fusion.  Depending on where the results of ATLAS and CMS
settle in these channels, it may or may not be possible to accommodate
the THDM interpretation. For this, more data is needed. 

We computed the signal cross section into diphotons and showed that 
it could explain the observations in the diphoton channel at the LHC for small values of
$\tan\beta$, as it can be seen in Figure~\ref{fig:1}. Given these
values of $\tan\beta$, we predict the branching fractions for 
the CP-even neutral states, shown in Figures~\ref{fig:2} and
\ref{fig:3}, and show that the new decay  modes $(h,H)\to AA$ and $(h,H)\to AZ$
are competitive enough so that these states, if heavy enough,   are
not affected by the LHC bounds on the SM Higgs. In general, it is
enough for CP-even masses to be above the $AA$ threshold, or $\simeq
250~$GeV,  for them not to be
bound by the existing LHC data up to rather very large values. The
confirmation of this scenario requires the search for these decay
modes in the $AA\to 4b's$, and $AZ\to \bar b b\ell^+\ell^-$~\cite{inprepa}.

The inverted THDM spectrum can be dynamically generated by models
where chiral fermions interact strongly with a new interaction at the
TeV scale, leading to their condensation and EWSB~\cite{fgthdm}. We
showed that the LHC diphoton signal can only  be accommodated in these
models if the new fermions are either colorless, so as to not affect
significantly the productions vertex, or in type~I scenarios which are
not motivated in fourth generation condensation models~\cite{fgthdm}. 

\vskip0.25in
\noindent{\bf Acknowledgments } 
G.~B. thanks Patrick Fox for helpful conversations and 
acknowledges the support of the Brazilian  National Council
for Technological and Scientific Development (CNPq). C.~E.~F.~H
thanks CAPES for its support.


\begin{thebibliography}{99}

\bibitem{atlas} The ATLAS Collaboration,  {\em Combination of Higgs
    Boson Searches with up to $4.9$ fb$^{-1}$ of pp Collision Data
    Taken at $\sqrt{s}=7~$TeV with the ATLAS Experiment at the LHC}, 
ATLAS-CONF-2011-163. 
%
\bibitem{cms}The CMS Collaboration, {\em Combination of CMS Searches
    for a Standard Model Higgs Boson}, CMS PAS HIG-11-032. 
%
\bibitem{atlasgg} G.~Aad {\it et al.}  [ATLAS Collaboration],
  %``Search for the Standard Model Higgs boson in the diphoton decay channel with 4.9 fb-1 of pp collisions at sqrt(s)=7 TeV with ATLAS,''
  Phys.\ Rev.\ Lett.\  {\bf 108}, 111803 (2012)
  [arXiv:1202.1414 [hep-ex]].
%
\bibitem{cmsgg}
 S.~Chatrchyan {\it et al.}  [CMS Collaboration],
  %``Search for the standard model Higgs boson decaying into two photons in pp collisions at sqrt(s)=7 TeV,''
  arXiv:1202.1487 [hep-ex].
%
\bibitem{atlaszz}
 G.~Aad {\it et al.}  [ATLAS Collaboration],
  %``Combined search for the Standard Model Higgs boson using up to 4.9 fb-1 of pp collision data at sqrt(s) = 7 TeV with the ATLAS detector at the LHC,''
  Phys.\ Lett.\ B {\bf 710}, 49 (2012)
  [arXiv:1202.1408 [hep-ex]].
%
\bibitem{cmszz}
 S.~Chatrchyan {\it et al.}  [CMS Collaboration],
  %``Search for the standard model Higgs boson in the decay channel H to ZZ to 4 leptons in pp collisions at sqrt(s) = 7 TeV,''
  arXiv:1202.1997 [hep-ex].
%
\bibitem{cmshjj}
CMS Collaboration, CMS-PAS-HIG-12-002.
%
\bibitem{atlashjj}
ATLAS Collaboration, ATLAS-CONF-2012-013.
%
%\bibitem{pdg}
%
 %K.~Nakamura {\it et al.}  [Particle Data Group Collaboration],
  %``Review of particle physics,''
 % J.\ Phys.\ G G {\bf 37}, 075021 (2010).
%
\bibitem{mssm}S.~P.~Martin,
  %``A Supersymmetry primer,''
  In *Kane, G.L. (ed.): Perspectives on supersymmetry II* 1-153
  [hep-ph/9709356].
%
\bibitem{maltoni} A  spectrum with $A$ the lightest is also
  considered in  S.~de Visscher, J.~-M.~Gerard, M.~Herquet, V.~Lemaitre, F.~Maltoni,
 % {\em Unconventional phenomenology of a minimal two-Higgs-doublet model},
  JHEP {\bf 0908}, 042 (2009).
  [arXiv:0904.0705 [hep-ph]].  Unlike here, a hierarchy in
  the heavier states is considered, and the phenomenology studied is
  different. 
%
\bibitem{sher} The implications of the data from ~\cite{atlas} and
  ~\cite{cms} for the normal THDM are studied in P.~M.~Ferreira, R.~Santos, M.~Sher and J.~P.~Silva,
  %``Implications of the LHC two-photon signal for two-Higgs-doublet models,''
  arXiv:1112.3277 [hep-ph]. It is assumed that the signals come
  from the lightest CP-even state, $h$.
%
\bibitem{strumia}
 P.~P.~Giardino, K.~Kannike, M.~Raidal and A.~Strumia,
  %``Reconstructing Higgs boson properties from the LHC and Tevatron data,''
  arXiv:1203.4254 [hep-ph].
%
\bibitem{fgthdm}  G.~Burdman and C.~E.~F.~Haluch,
  %``Two Higgs Doublets from Fermion Condensation,''
  JHEP {\bf 1112}, 038 (2011)
  [arXiv:1109.3914 [hep-ph]].
%
\bibitem{barger} V.~D.~Barger, J.~L.~Hewett and R.~J.~N.~Phillips,
  %``New Constraints On The Charged Higgs Sector In Two Higgs Doublet Models,''
  Phys.\ Rev.\ D {\bf 41}, 3421 (1990).
%
\bibitem{brancosher} For a recent review see 
 G.~C.~Branco, P.~M.~Ferreira, L.~Lavoura, M.~N.~Rebelo, M.~Sher and J.~P.~Silva,
  %``Theory and phenomenology of two-Higgs-doublet models,''
  arXiv:1106.0034 [hep-ph].
%
\bibitem{spira}  M.~Spira, A.~Djouadi, D.~Graudenz and P.~M.~Zerwas,
  %``Higgs boson production at the LHC,''
  Nucl.\ Phys.\ B {\bf 453}, 17 (1995)
  [hep-ph/9504378].
%
\bibitem{flavorbounds} M.~Misiak, H.~M.~Asatrian, K.~Bieri, M.~Czakon, A.~Czarnecki, T.~Ewerth, A.~Ferroglia and P.~Gambino {\it et al.},
  %``Estimate of B(anti-B ---> X(s) gamma) at O(alpha(s)**2),''
  Phys.\ Rev.\ Lett.\  {\bf 98}, 022002 (2007)
  [hep-ph/0609232].
See also O.~Deschamps, S.~Descotes-Genon, S.~Monteil, V.~Niess, S.~T'Jampens and V.~Tisserand,
 %{\em The Two Higgs Doublet of Type II facing flavour physics data}, 
  Phys.\ Rev.\  D {\bf 82}, 073012 (2010)
  [arXiv:0907.5135 [hep-ph]].
%
\bibitem{inprepa} G.~Burdman, C.~E.~F.~Haluch and R.~D.~Matheus, in
  preparation. 
%
\bibitem{pdg}K.~Nakamura {\it et al.}  [Particle Data Group Collaboration],
  %``Review of particle physics,''
  J.\ Phys.\ G G {\bf 37}, 075021 (2010).
%
\bibitem{tlunitarity} We check tree-level unitarity using the 2HDM
  calculator from D.~Eriksson, J.~Rathsman and O.~Stal,
  %``2HDMC: Two-Higgs-Doublet Model Calculator Physics and Manual,''
  Comput.\ Phys.\ Commun.\  {\bf 181}, 189 (2010)
  [arXiv:0902.0851 [hep-ph]];   D.~Eriksson, J.~Rathsman and O.~Stal,
  %``2HDMC: Two-Higgs-doublet model calculator,''
  Comput.\ Phys.\ Commun.\  {\bf 181}, 833 (2010).
% 
\bibitem{nmssm} 
R.~Dermisek, J.~F.~Gunion,
 % {\em New constraints on a light CP-odd Higgs boson and related NMSSM Ideal Higgs Scenarios},
  Phys.\ Rev.\  {\bf D81}, 075003 (2010).
  [arXiv:1002.1971 [hep-ph]], and references therein.
% 
\bibitem{sekhar} 
  R.~S.~Chivukula, P.~Ittisamai, E.~H.~Simmons and J.~Ren,
  %``Technipion Limits from LHC Higgs Searches,''
  Phys.\ Rev.\ D {\bf 84}, 115025 (2011)
  [arXiv:1110.3688 [hep-ph]].
%
\bibitem{fg1} G.~Burdman and L.~Da Rold,
%  {\em Electroweak Symmetry Breaking from a Holographic Fourth   Generation}, 
  JHEP {\bf 0712}, 086 (2007)
  [arXiv:0710.0623 [hep-ph]].
%
\bibitem{he}  
 C.~Anastasiou, S.~Buehler, E.~Furlan, F.~Herzog and A.~Lazopoulos,
  %``Higgs production cross-section in a Standard Model with four generations at the LHC,''
  Phys.\ Lett.\ B {\bf 702}, 224 (2011)
  [arXiv:1103.3645 [hep-ph]].
 A.~Denner, S.~Dittmaier, A.~Muck, G.~Passarino, M.~Spira, C.~Sturm, S.~Uccirati and M.~M.~Weber,
  %``Higgs production and decay with a fourth Standard-Model-like fermion generation,''
  arXiv:1111.6395 [hep-ph].
 G.~Guo, B.~Ren and X.~-G.~He,
  %``LHC Evidence Of A 126 GeV Higgs Boson From $H \to \gamma \gamma$ With Three And Four Generations,''
  arXiv:1112.3188 [hep-ph].
\end{thebibliography}
\end{document}